\documentclass[aps,prl,twocolumn]{revtex4-2}
\pdfoutput=1
\usepackage{graphicx}
\usepackage{grffile} %needed when image file name has multiple dots
\usepackage{dcolumn}
\usepackage{bm,amssymb,amsmath}
\usepackage[pdftex]{hyperref}
\hypersetup{colorlinks=true,citecolor=blue,linkcolor=red,urlcolor=blue}
\usepackage[all]{hypcap}
\usepackage{color}
\definecolor{red}{rgb}{0.8, 0.0, 0.0}
\definecolor{blue}{rgb}{0.06, 0.2, 0.65}
\definecolor{green}{rgb}{0,0.6,0}

\begin{document}

\title{Growth and prediction of plastic strain in metallic glasses}
\author{Tero Mäkinen$^{a,*1}$, Anshul D. S. Parmar$^{b,*2}$, Silvia Bonfanti$^{b,c}$, Mikko Alava$^{a,b}$}
\affiliation{
    $^a$ Aalto University, Department of Applied Physics, PO Box 11000, 00076 Aalto, Espoo, Finland\\
    $^b$ NOMATEN Centre of Excellence, National Center for Nuclear Research, ul. A. Soltana 7, 05-400 Swierk/Otwock, Poland\\
    $^c$ Center for Complexity and Biosystems, Department of Physics `Aldo Pontremoli', University of Milan, Milano, Italy
}
\email{T.M. and A.D.S.P contributed equally to this work.
To whom correspondence should be addressed. E-mails: $^1$tero.j.makinen@aalto.fi and $^2$anshul.parmar@ncbj.gov.pl}

\begin{abstract}
Predicting the failure and plasticity of solids remains a longstanding challenge, with broad implications for materials design and functional reliability. 
Disordered solids like metallic glasses can fail either abruptly or gradually without clear precursors, and the mechanical response depends strongly on composition, thermal history and deformation protocol---impeding generalizable modeling.
While deep learning methods offer predictive power, they often rely on numerous input parameters, hindering interpretability, methodology advancement and practical deployment.
Here, we propose a macroscopic, physically grounded approach that uses plastic strain accumulation in the elastic regime to robustly predict deformation and yield. This method reduces complexity and improves interpretability, offering a practical alternative for disordered materials.
For the Cu-Zr-(Al) metallic glasses prepared with varied annealing,
we identify two limiting regimes of plastic strain growth: power-law in poorly annealed and exponential in well-annealed samples.
A physics-informed framework with Bayesian inference extracts growth parameters from stress-strain data within $\sim$5\% strain, enabling early prediction of bulk response and yield point, well before the failure.
The predictive performance improves with annealing, and bulk plasticity correlates with the microscopic plastic activity from scattered to growth near yielding. 
This work presents a physically interpretable and experimentally relevant framework 
for predicting plasticity and failure in metallic glasses from early mechanical response, offering both theoretical insights and practical tools for material characterization and design.
\end{abstract}

\maketitle

Mechanical response in solids progresses from an initial elastic regime to plastic deformation, culminating in yielding and structural failure~\cite{berthier2025yielding}. Predicting plastic response---identifying micro/macroscopic precursors and determining whether failure to be gradual (ductile) or abrupt (brittle)---remains a central challenge in materials science and engineering~\cite{mises1913mechanik, Hill1948, divoux2024ductile}. In crystalline solids, prediction of mechanical response is partially addressed by the framework of crystal plasticity~\cite{Aravas1991, Anand1996}, though the complete nature of defect dynamics and interactions remains elusive~\cite{miller2008nonlocal,miguel2001intermittent,salmenjoki2020plastic,dahmen2009micromechanical,ispanovity2010submicron,ispanovity2014avalanches,salmenjoki2018machine}. Amorphous materials---such as metallic glasses, molecular, and colloidal systems---represent the extreme limit of structural disorder.  
Unlike crystals, disordered solids lack apparent structural precursors---compounded by sensitivity to sample preparation and loading protocols~\cite{leishangthem2017yielding,ozawa2018random,parmar2019strain,bhaumik2021role,yeh2020glass}, which makes predicting mechanical response and failure extremely challenging~\cite{spaepen1977microscopic}. 
Overcoming these obstacles would enable accurate predictions of yielding and plastic flow in disordered systems, with significant implications for both fundamental understanding and practical applications.

Macroscale plasticity arises from localized shear events in shear transformation zones (STZs)~\cite{argon1979plastic,falk1998dynamics,nicolas2018deformation}, which are influenced by atomic-scale heterogeneity. Computational simulations have qualitatively examined various structural features---such as locally favored packings~\cite{Coslovich2007,Paret2020}, compactness~\cite{tong2018revealing}, local yield stress and shear modulus~\cite{Tsamados2009,patinet2016connecting}, soft vibrational modes~\cite{WidmerCooper2008,Tanguy2010,Manning2011,Gartner2016,bonfanti2019elementary}, and underlying topological order~\cite{wu2023topology,wu2024geometry}---all of which affect how plasticity emerges under deformation~\cite{richard2020predicting}. Modern machine learning (ML) methods~\cite{jung2025roadmap} offer a promising pathway to uncover complex structural signatures that are otherwise obscure. Recent studies have proposed various ML techniques---including support vector machines (SVMs)~\cite{Schoenholz2016}, graph neural networks (GNNs)~\cite{Bapst2020}, unsupervised learning~\cite{Boattini2019,Boattini2020}---as well as elastoplastic models~\cite{zhang2021interplay,xu2024stochastic} to efficiently reveal underlying structure-property relationships.
In the context of structure-mechanical response, ML and deep learning approaches have been used to build predictive frameworks---for example, convolutional neural networks (CNNs) to predict local plastic rearrangements~\cite{fan2021predicting}, physics-informed networks to infer atomic structure~\cite{bodker2022predicting}, neural networks for prediction and nature of failure~\cite{font2022predicting,fan2022predicting}, and GNN-based representations of microstructural deformation~\cite{thomas2023materials,zhai2025stress}. 
As defect evolution and interaction depend strongly on sample preparation and deformation protocols~\cite{leishangthem2017yielding,ozawa2018random}, no universal structural signature currently exists for predicting plasticity and catastrophic failure~\cite{richard2020predicting}. While deep learning effectively identifies structural defects and affinity with plastic events, its predictions---especially concerning macroscopic yield---often lack interpretability and physical insight.

\begin{figure*}[tb]
\centering
\includegraphics[width=\textwidth]{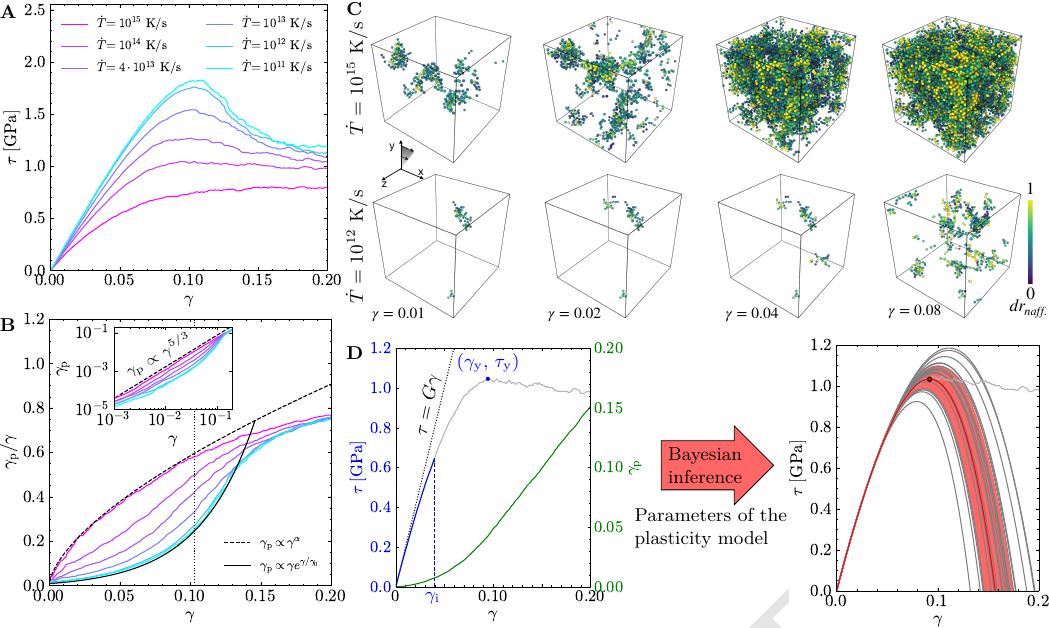}
\caption{\textbf{A}~The stress-strain curves for glasses prepared with different cooling rates~$\dot{T}$.
\textbf{B}~The proportion of plastic strain~$\gamma_{\rm p}$ as a function of strain~$\gamma$ for each cooling rate (colors as in panel A). The dashed and solid black lines are are fits (up to $\gamma = 0.10$) to $\gamma_{\rm p} \propto \gamma^\alpha$ and $\gamma_{\rm p} \propto \gamma e^{\gamma/\gamma_0}$ for the extreme cooling rate cases, vertically shifted by 0.01 for clarity. The vertical dotted line corresponds to the average yield strain.
The inset shows the plastic strain in a log-log plot, along with a power-law behavior (black dashed line).
\textbf{C}~The particles are colored by the non-affine displacement $dr_{\rm naff.}$ under a stress-drop, shown cumulatively for two cooling rates (rows) and four different global strains below the yield (columns).
\textbf{D}~The Bayesian prediction scheme: by reading the stress-strain curve up to a strain~$\gamma_{\rm i}$ (blue part) we predict the yield point $(\gamma_{\rm y}, \, \tau_{\rm y})$. This is done based on the accumulated plastic strain~$\gamma_{\rm p}$ (green), which is determined from the difference to the perfect elastic behavior (black dashed line). We then infer the parameter distributions for the plasticity model, generate samples (each corresponding to a stress strain curve shown in grey), and finally average them to get the prediction (red line) and uncertainty estimates (light red shaded area). From the prediction we can get the yield point (red point).
}\label{fig:fig1}
\end{figure*}

In this work, we introduce a robust Bayesian machine learning framework to predict yielding in metallic glasses using { {macroscopic}} 
plasticity features.
Our approach focuses on plastic strain accumulation within the elastic regime—a physically meaningful and experimentally accessible indicator—to accurately predict yield stress and strain (see Fig.~\ref{fig:fig1}), unlike conventional models that rely on numerous parameters or detailed structural data. 
This macroscopic signal correlates with local rearrangements, such as STZs, underscoring its physical relevance.
We demonstrate the broad applicability of the framework 
across ductile materials and preparation conditions, without requiring detailed microstructural, compositional, or thermal history inputs—making it both practical and interpretable.

\section*{Simulation and model details}
The data presented in this study are primarily based on simulations of an equiatomic Cu-Zr metallic glass, as described in Ref.~\cite{makinen2025avalanches}. Additional results for Cu-Zr(-Al) compositions are provided in the Supplementary Information (SI). 
Each system contains $N = 6000$ particles (see SI for a discussion on system size effects), and six different cooling rates~($\dot{T}$), ranging from $10^{11}$ to $10^{15}$~K/s, are employed to generate glasses with varying degrees of annealing, form well to poorly annealed states. 
The cooling process is carried out using a hybrid molecular dynamics–Monte Carlo scheme~\cite{sadigh2012scalable,alvarez2023simulated,zhang2022shear}, followed by shear deformation simulations using the athermal quasistatic loading protocol (AQS, see Materials and Methods for details).
To obtain more experimental-representative stress–strain responses, we average the individual curves over 100 independent realizations of the disordered structure (50 for the slowest cooling rate). This statistical averaging approximates the response of a macroscopic sample composed of many representative volume elements.

\begin{figure*}[tb]
    \centering
    \includegraphics[width=\textwidth]{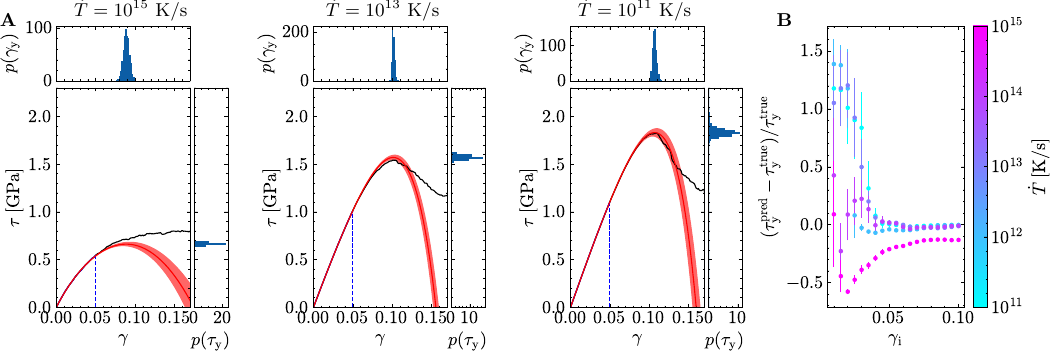}
    \caption{
    \textbf{A}~The Bayesian prediction of the stress-strain curve for three different cooling rates~$\dot{T}$. The red curve is the average stress-strain prediction (shaded area corresponding to the uncertainty estimate), made based on the data before $\gamma_{\rm i} = 5$~\% (blue dashed line). The probability distributions of the yield stress~$p(\tau_{\rm y})$ and yield strain~$p(\gamma_{\rm y})$ derived from the individual prediction samples are shown on top and on the side of the figures.
    \textbf{B}~The relative difference between the predicted~$\tau_{\rm y}^{\rm pred}$ and true value~$\tau_{\rm y}^{\rm true}$ of the yield stress as a function of the prediction strain~$\gamma_{\rm i}$ for the different cooling rates. Although the yield stress is ill-defined in poorly annealed glasses, the predictions remain sufficiently accurate (with $\sim$20\% underestimation of stress).}
    \label{fig:predicted_ss}
\end{figure*}

\section*{Results} 
\subsection*{Mechanical response} 
We obtain the mechanical response of the samples by shearing them using the AQS protocol across the yielding strain and record the stress-strain response (see Fig.~\ref{fig:fig1}A). Depending on the cooling rate~$\dot{T}$ used for glass preparation, we observe strongly varying behavior. For slow cooling rates, we observe a long linear---close to elastic---behavior extending to around $\gamma = 5$~\%. For faster cooling rates, the initial slope of the stress-strain curves, i.e., shear moduli $G$, is lower, and the variation from linear behavior starts almost immediately.
Additionally, slower cooling rates result in higher peak stresses, with a more pronounced maximum in the stress–strain curves. We define the maximum shear stress as the yield stress~$\tau_{\rm y}$, and the corresponding strain as the yield strain~$\gamma_{\rm y}$.
This working definition may be ambiguous for poorly annealed glasses, where the maximum-stress 
can occur within the steady-state flow regime. Nevertheless, we adopt this single definition of yield 
to ensure consistent comparison across glasses prepared with different annealing (see SI for an alternative yield strain definition).

The initial modulus~$G$ allows for the determination of the plastic strain using the relation $\gamma_{\rm p} = \gamma - \frac{\tau}{G}$ (see SI for discussion on the constancy of $G$). Looking at the evolution of this quantity (illustrated in Fig.~\ref{fig:fig1}B), we observe two extreme cases: a power-law $\gamma_{\rm p} \propto \gamma^\alpha$ for fast cooling rates and and a mix of linear and exponential behavior $\gamma_{\rm p} \propto \gamma e^{\gamma/\gamma_0}$ for the slow cooling rates. The latter is essentially exponential behavior, with a linear part at low strains $\gamma \ll \gamma_0$---giving the correct $\gamma_{\rm p} \to 0$ behavior at the $\gamma \to 0$ limit.

Figure~\ref{fig:fig1}C provides a microscopic visualization of the accumulated plastic strain in glasses annealed at cooling rates $\dot{T} = 10^{15}$ (top) and $\dot{T} = 10^{12}$ (bottom), for systems with $N = 24000$ particles. Particles 
rearranged in a plastic event are highlighted based on a non-affine displacement threshold of $dr_{\text{naff.}} > 0.3~\text{\AA}$ (see Materials and Methods and Ref.~\cite{makinen2025avalanches}). These highlighted particles, identified during individual plastic events and shown cumulatively over a range of strain values, represent the evolving spatial profile of local plastic activity, which can be considered representative of shear transformation zones or softness~\cite{richard2020predicting}. 

Plastic activity appears spatially extensive in the poorly annealed glass, whereas in the well-annealed glass it remains more localized. This localization, understood to be correlated with the spatial distribution of soft regions, is also observed in the local yield stress and elastic moduli~\cite{Tsamados2009,patinet2016connecting,richard2020predicting}. As yielding approaches, we observe that 
these initially isolated soft regions grow, leading to relative growth of the clusters of plastically deformed particles---consistent with the observed growth behavior of the global plastic strain.

\subsection*{Bayesian prediction model}
To predict the yielding behavior of the glasses we construct a function that can encompass both of the extreme cases and interpolate between them, namely
\begin{equation} \label{eq:plastic_strain}
    \gamma_{\rm p} = A \gamma^\alpha e^{\gamma/\gamma_0},
\end{equation}
where we explictly introduce the prefactor $A$. This reproduces the slow cooling rate near-exponential behavior with $\alpha = 1$, and in the $\gamma_0 \to \infty$ limit depicts the pure power-law behavior, as observed for the faster cooling. 
Using this function, we can make early predictions of the yield point $(\gamma_{\rm y}, \, \tau_{\rm y})$, based on the stress-strain data read only up to a low strain value~$\gamma_{\rm i}$.
The goal is to infer the values of the parameter set $\theta = \{ A, \gamma_0, \alpha \}$ based on the observed data $\mathcal{D}$ (stress-strain data up to strain~$\gamma_{\rm i}$) and to extrapolate the stress-strain behavior at future strains from $\tau = G (\gamma - \gamma_{\rm p})$. This process is shown in Fig.~\ref{fig:fig1}D.

We do prediction using Bayesian inference, which starts from the Bayes' theorem
\begin{equation} \label{eq:bayes}
    p(\theta | \mathcal{D}) = \frac{p(\mathcal{D} | \theta) p(\theta)}{p(\mathcal{D})}
    = \frac{p(\mathcal{D} | \theta) p(\theta)}{\int p(\mathcal{D} | \theta) p(\theta) \mathrm{d} \theta},
\end{equation}
where $p(\theta | \mathcal{D})$ is the posterior distribution of the parameters conditional on the observed data, $p(\mathcal{D} | \theta)$ the likelihood, and $p(\theta)$ the prior distribution encoding our knowledge about the parameters before observing any data. The integral over all the possible parameter values in the marginal likelihood $p(\mathcal{D})$ is generally analytically intractable, but Markov chain Monte Carlo~(MCMC) methods can be used to generate a sequence of parameter values, the distribution of which approximates the posterior distribution.
Compared to simple curve fitting, Bayesian inference provides both regularization through priors and robust uncertainty quantification, which is used to assess the reliability of predictions and the convergence of the ML algorithm.
Here we read the stress-strain data $(\gamma, \tau)$ up to a strain $\gamma_{\rm i}$ and use this as the observed data~$\mathcal{D}$. Using MCMC 
we then sample from the posterior distribution (Eq.~\ref{eq:bayes}), and for each sample can compute the plastic strain~(Eq.~\ref{eq:plastic_strain}), generate a stress-strain curve, and finally determine the corresponding yield point $(\gamma_{\rm y}, \tau_{\rm y})$. See Materials and Methods section for more details on the implementation.\\

Taking as an example the prediction strain of $\gamma_{\rm i} = 5$~\% shows the power of the proposed framework %approach 
(see Fig.~\ref{fig:predicted_ss}A). By only considering small strains, in the near-linear elastic regime, we still observe enough plasticity to be able to infer fairly narrow distributions for the model parameters~$\theta$, thus having fairly narrow distributions for the predictions of yield stress~$\tau_{\rm y}$ and strain~$\gamma_{\rm y}$.
For the slower cooling rates the prediction is excellent, closely following the shape of the stress-strain curves up to yielding. 
For poorly annealed glasses, where the yield point is ill-defined and plastic strain accumulates slower than predicted, the model tends to underestimate the yield stress by around 20~\%. Nevertheless, the predictions remain reasonably accurate and relevant for practical engineering applications (see SI for further discussion).
The prediction accuracy naturally increases, as one takes more data into account (i.e. increases $\gamma_{\rm i}$) and the predictions start to converge to the correct values at around $\gamma_{\rm i} = 5$~\% (see Fig.~\ref{fig:predicted_ss}B).
See SI for more discussion on the posterior distributions of the parameters (including the connection between $A$ and $\gamma_0$), as well as the posterior distributions of $\tau_{\rm y}$ and $\gamma_{\rm y}$, with changing $\gamma_{\rm i}$.\\

\begin{figure}[tb!]
    \centering
    \includegraphics[width=\columnwidth]{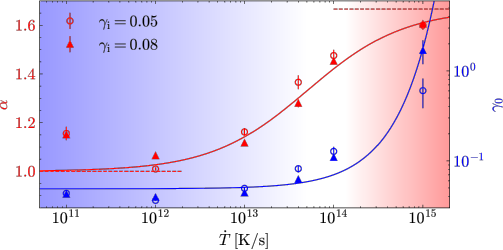}
    \caption{The mean behavior of the inferred values of the parameters of Eq.~\ref{eq:plastic_strain} for $\alpha$ (red) and $\gamma_0$ (blue) for each cooling rate, and for two different Bayesian prediction strains~$\gamma_{\rm i}$. The dashed red lines correspond to the linear ($\alpha = 1$) and percolation cluster ($\alpha = 5/3$) behaviors, and the solid red line a logistic fit going from one limiting behavior into the other.
    The solid blue line corresponds to an exponential fit. 
    The background color indicates the cooling rates dominated by the exponential behavior~(blue) and the ones dominated by the power-law behavior~(red). Both the fits and color gradient serve as guides to the eye.}
    \label{fig:means}
\end{figure}

\subsection*{Two limiting behaviors}
The inferred posterior distributions of the parameters of $\alpha$ and $\gamma_0$ of Eq.~\ref{eq:plastic_strain} are very narrow (see errorbars in Fig.~\ref{fig:means}). 
We can also see that the posterior distributions do not evolve much going from strain $\gamma_{\rm i} = 0.05$ to $\gamma_{\rm i} = 0.08$ (see difference of open circles and filled triangles in Fig.~\ref{fig:means}).
One can also see the expected behavior for the extreme cases: for slow cooling rates $\gamma_0$ is low, indicating exponential behavior with the low strain power-law regime having close to linear behavior ($\alpha \approx 1$), and for high cooling rates $\gamma_0$ is very high, indicating dominant power-law behavior. This is qualitatively shown in the background color of Fig.~\ref{fig:means}.
Such power-law and exponential behaviors have previously been seen also in dislocation plasticity~\cite{ispanovity2010submicron}.

We next outline an argument for the development of plasticity in ``bad" glasses where the $\gamma_{\rm p}$ exhibits a power-law like growth. With the observation that the number of atoms participating in yielding scales as $n \simeq \gamma^\alpha$, we write the growth law for $n$ as 
\begin{equation}
    \frac{\partial n }{\partial \gamma} \simeq n^{(\alpha -1)/\alpha}.
\end{equation}
We next assume that the growth takes place in diffuse clusters of internal dimension $d_{\rm f}$ and a surface dimension $d_{\rm s}$ for $n$ and cluster size~$S$ as a function of the cluster linear dimension $R$. If the growth rate $\Delta n$ is taken proportional to $S$, we solve $\alpha$ and obtain that
\begin{equation}
    \alpha = \frac{d_{\rm f}}{d_{\rm f} - d_{\rm s}} .
\end{equation}
For 3D isotropic percolation these exponents have the values $d_{\rm f} \simeq 2.53$ and $d_{\rm s} \simeq 1.03$ which gives with quite a good approximation $\alpha = 5/3$. This matches the limiting behavior we see at high $\dot{T}$ (Fig.~\ref{fig:means}).
The microscopic observations for the poorly annealed glass are also consistent with the fractal dimension of 3D isotropic percolation, which is observed well beyond the yield point and  transient 
shear band (see SI).
These limits should be ``universal" in bounding the plasticity development, thus the shape of the stress-strain curve, and lead to the demonstration in the SI how various compositions of the CuZrAl family at a fixed cooling rate map into CuZr with different cooling rates.

\begin{figure}[t!]
    \centering
    \includegraphics[width=\columnwidth]{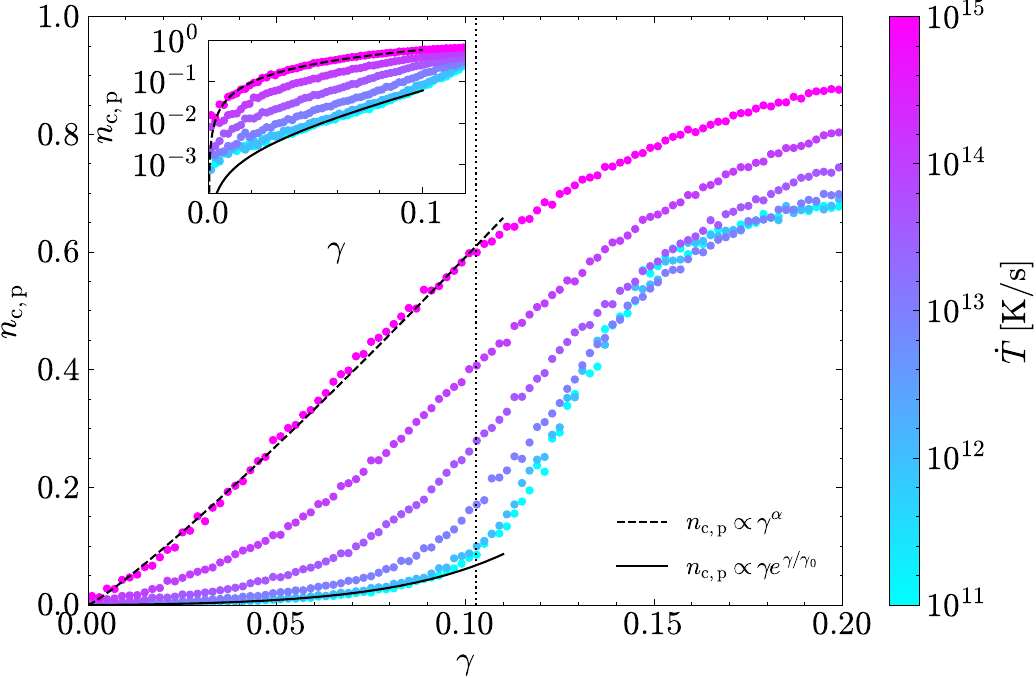}
    \caption{
    The cumulative fraction of yielded particles $n_{\rm c, p}$ as a function of the applied strain $\gamma$ for the different cooling rates. The inset shows the same data in a logarithmic-linear plot to illustrate the exponential behavior in the plasticity in the pre-yield regime. The vertical dashed line marks the averaged yielding strain for all the cooling rates.
    }
    \label{fig:plasticity}
\end{figure}

\subsection*{Microscopic nature of plasticity}
To gain microscopic insights with a structural perspective, we examine the fraction of particles ($n_{p}$) involved in irreversible plastic deformation. During an avalanche, a particle is considered part of the plastic core if its non-affine displacement exceeds 0.3~\AA. This displacement threshold allows the identification of particles undergoing plastic rearrangements at the microscopic level (see Materials and Methods for details and Ref.~\cite{leishangthem2017yielding,makinen2025avalanches}). Recognizing such particles serves as a tool for identifying local rearrangments (STZs), the fundamental units of plastic deformation. The particles involved in an avalanche exhibit a physically consistent correlation with the active participation in the bulk stress drops (as discussed in SI and Ref.~\cite{makinen2025avalanches}), further asserting the physical relevance of this observation.

To track the development of plasticity, we monitor these particle-fractions cumulatively ($n_{\rm c,p}$)---from the initial undeformed state to well beyond the yield strain. Fig.~\ref{fig:plasticity} shows the cumulative fraction of particles exhibiting plastic activity for various cooling rates. 
In the nearly elastic regime, the poorly annealed glass exhibits weak power-law behavior with an exponent of $\alpha = 1.12$. The observed exponent is noticeably smaller than that for bulk plastic strain, possibly due to the absence of long-range elastoplastic responses in the present microscopic analysis, which are inherently difficult to distinguish from local effects~\cite{leishangthem2017yielding}. As the degree of annealing increases, the evolution of plasticity gradually transitions toward an exponential form. This cumulative microscopic evolution resembles the behavior of the macroscopic plastic strain, highlighting a consistent correlation between atomistic dynamics and bulk mechanical response.

\begin{figure}[t!]
    \centering
    \includegraphics[width=\columnwidth]{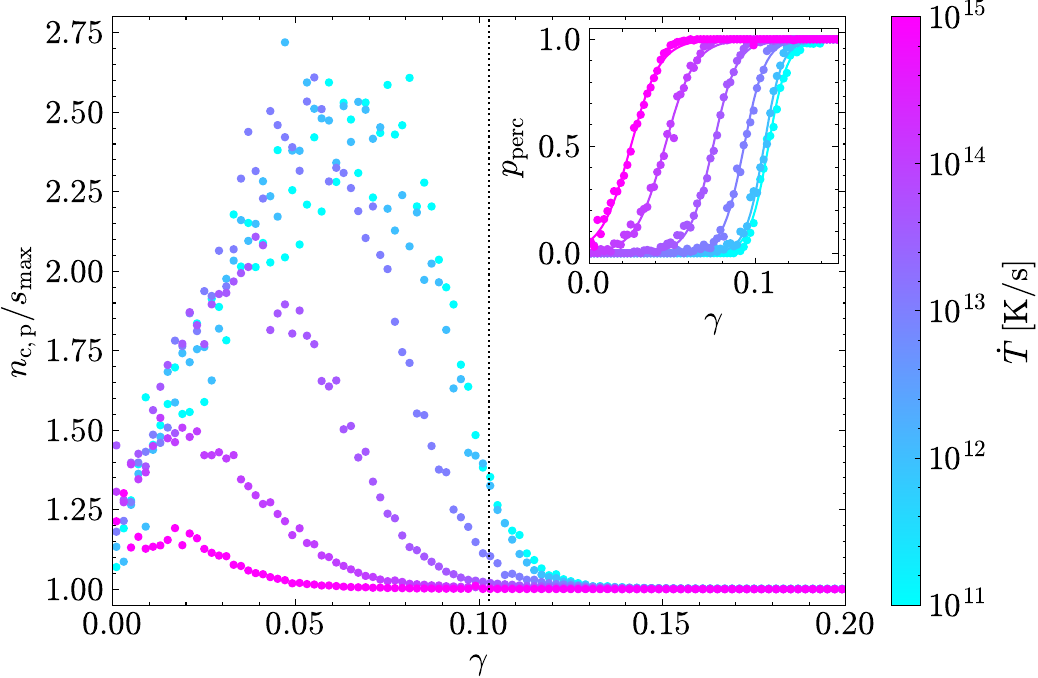}
    \caption{
    The ratio between the cumulative fraction of yielded particles $n_{\rm c, p}$ and the size of maximum cluster $s_{\rm max}$ as a function of the applied strain $\gamma$ for the different cooling rates. In the pre-yield regime, the systematic increase in the peak of the $n_{\rm c,p} / s_{\rm max}$ suggests that plastic events are distinct till the strain $\gamma\sim 0.05$, which further percolates to form the spanning-cluster.  The inset shows the percolation probability $p_{\rm perc}$ and a logistic fit.
    The vertical dashed line marks the averaged yielding strain for all the cooling rates. 
    }\label{fig:cluster}
\end{figure}

\subsection*{Microscopic signal for machine learning predictability}
To further explore the development of bulk plasticity under shear, we perform a spatial analysis based on the cumulative clustering of particles undergoing plastic rearrangements. Two particles are considered part of the same cluster if they are neighbors, defined as being within the first coordination shell according to the radial distribution function. Similar to the cumulative fraction of yielded particles ($n_{\rm c,p}$) and bulk plastic strain, the size of the largest cluster also increases and exhibits a comparable dependence on the cooling rate (see SI).  
 
To highlight the spatial nature of plasticity, Fig.~\ref{fig:cluster} shows the ratio of the cumulative fraction of particles undergoing plastic rearrangements, $n_{c,\mathrm{p}}$, to the size of the largest cluster, $s_{\mathrm{max}}$, as a function of strain $\gamma$ across cooling rates. As noted in Fig.~\ref{fig:fig1}, plastic activity is spatially extensive in poorly annealed glasses, forming percolating clusters early in deformation. In contrast, plasticity in better-annealed glasses is more localized: the ratio gradually increases and peaks, which depends on the cooling rate, indicating that plastic events remain spatially distinct within the initial-elastic limit (up to 5~\% strain). Beyond this point, as observed, events begin to merge into a spanning cluster and grow. However, full percolation requires additional loading and further distinct plastic events.
These findings demonstrate that the spatial distribution and growth of plasticity, even in the near-elastic regime, encode structurally relevant information, observed to be effectively captured by our ML-based model and motivating its extension to brittle glasses to test broader applicability.

\section*{Conclusions}

This work presents a predictive framework for the mechanical response of Cu-Zr(-Al) metallic glasses under shear, combining physical insight with Bayesian machine learning. By analyzing plastic strain evolution in the early elastic regime, we show that yield stress and strain can be accurately predicted across a broad range of cooling rates associated with ductile failure. Plasticity develops progressively and reflects the thermal history of the glass, with pre-yield activity more localized in well-annealed samples and more extensive in poorly annealed ones. Bulk plastic strain, associated with the plastic rearrangements, emerges as a key predictive feature.

Our approach replaces the complexity of multi-variable models into a small set of physically meaningful, global observables, e.g., stress and plastic strain, without relying on detailed microscopic inputs that are often specific to a given system or thermal history~\cite{richard2020predicting}. It is computationally efficient, broadly applicable across different preparation conditions, compositions (see SI), and physically interpretable. Importantly, our method overcomes key limitations of commonly used deep learning approaches, offering a more transparent, transferable framework. While demonstrated under athermal quasi-static shear conditions, extending this work to finite temperatures, shear rates, and ductile-to-brittle transitions offers a promising direction for future work~\cite{ozawa2018random,singh2020brittle}. Our findings encourage a deeper exploration of plasticity growth with multiscale modeling and its connection to yielding through theoretical studies, atomistic simulations, and mesoscale elastoplastic models.
This study shows that physics-informed machine learning with bulk-plasticity can provide robust, interpretable, and transferable predictions of yielding in disordered materials, directly relevant to experiments and applications.

\section*{Materials and Methods}

\subsection*{System} 
In this study, we investigate the mechanical response of metallic glasses using the binary Cu$_{0.50}$Zr$_{0.50}$ and a wide range of ternary Cu-Zr-Al amorphous compositions. %Cu$_{0.46}$Zr$_{0.46}$Al$_{0.08}$. 
Molecular dynamics (MD) simulations are performed using LAMMPS~\cite{LAMMPS}, with atomic interactions described by an embedded atom method (EAM) interatomic potential~\cite{cheng2009atomic}. Simulations are primarily conducted on a system of 6000 atoms contained in a cubic box with periodic boundary conditions. To evaluate system size effects, larger systems comprising up to $N = 24000$ atoms are also simulated.

\subsection*{Sample preparation}
Glass preparation is performed using a hybrid Molecular Dynamics-Monte Carlo (MD+MC) scheme within the variance-constrained semi-grand canonical (VC-SGC) ensemble~\cite{sadigh2012scalable}. Rather than swapping atom types between a pair of particles, the method randomly selects a single atom and attempts to change its type. The acceptance of this trial move follows the Metropolis criterion, ensuring detailed balance while maintaining the target composition and efficient sampling of configurational space. To minimize compositional deviations, the parameters for chemical potential to perform the hybrid MD+MC simulations under the VC-SGC ensemble are provided by Ref.~\cite{alvarez2023simulated}.
A similar hybrid scheme has also previously been applied to Cu-Zr systems~\cite{zhang2022shear}.

The quenching is performed starting from molten metals at a high temperature well above the melting point using a fixed cooling rate~$\dot{T}$ at fixed pressure $P=0$~bar~\cite{alvarez2023simulated}. In the MD+MC~scheme a MC~cycle consisting of $N$ attempts is performed every 20~MD~steps. The independent liquid samples are cooled from 2000~K to 300~K, with cooling rates ranging from $10^{11}$ to $10^{15}$, ensuring an extended range of supercooled states~\cite{makinen2025avalanches}.

\subsection*{Mechanical deformation} 
We perform athermal quasi-static shear (AQS) simulations under Lees–Edwards periodic boundary conditions. The simulation cell is incrementally sheared in the $xy$-plane with a strain step of $\delta\gamma = 5\times10^{-6}$, followed by energy minimization using the conjugate gradient method. This protocol mimics the experimentally relevant conditions of low temperature and low shear rate, enabling a consistent stress–strain response for mechanical characterization. For optimized computation and structural analysis in the $N=24000$ systems, we deformed the system using strain increments of $\delta\gamma = 10^{-4}$. To improve statistical averaging of avalanches, shear was applied independently along the $xy$, $yz$, and $zx$ planes, since the system is isotropic. 

To quantify the mechanical response, we focus on the shear modulus ($G$) and the yield stress ($\tau_{\rm y}$).
The shear modulus is determined from the stress-strain curves by averaging the positive slopes for the first 0.5~\% of strain in each of the independent realizations of the atomic configurations.
The yield stress is determined from the average stress–strain curve over all the realizations and defined as the maximum shear stress attained during deformation. See Fig.~\ref{fig:fig1}D for an illustration of the stress–strain curve analysis.

\subsection*{Bayesian inference}
The MCMC sampling from the posterior distribution (Eq.~\ref{eq:bayes}) is done using PyMC~\cite{abril2023pymc} software and the No-U-Turn sampler~(NUTS)~\cite{hoffman2014no}. We run 4~chains in parallel and for each discard the first 1000 tuning samples, after which we record 2000 samples.
The full model is 
\begin{equation}
    \tau = G \left(\gamma - A \gamma^\alpha e^{\gamma/\gamma_0} \right) + \epsilon
\end{equation}
where $G$ is the constant shear modulus (see SI for discussion on the constancy of the modulus) and $\epsilon$ is the noise observed.
The noise is implemented through the use of a Gaussian likelihood on the logarithm of the stress
\begin{equation}
    p(\mathcal{D} | \theta) = \prod_k \frac{\exp\left( - \frac{\left( \log \tau_k - \log \left[ G \left(\gamma_k - A \gamma_k^\alpha e^{\frac{\gamma_k}{\gamma_0}} \right) \right] \right)^2}{2 \sigma_\epsilon^2} \right)}{\sqrt{2 \pi \sigma_\epsilon^2}}
\end{equation}
where the index $k$ runs through all the datapoints in $\mathcal{D}$. 
The logarithm is there to give more focus to the early part of the stress-strain curves, where the stress is low.
We have used a half-normal prior for $\sigma_\epsilon$ with a standard deviation of 0.1~GPa.
Based on the knowledge of the valid range for the exponent $\alpha$, we use an informative prior, namely the uniform distribution between 1 and 2. For the parameters $A$ and $\gamma_0$ we use weakly informative lognormal priors, with the probability density function $p(x) \propto \exp\left[ - (\ln x - \mu)^2 / (2 \sigma^2) \right]$, and the values ($\mu = \ln(0.2)$, $\sigma = 4$) for both $A$ and $\gamma_0$.

Each sample of $A$, $\alpha$, and $\gamma_0$ can then be used to generate the evolution of the plastic strain using Eq.~\ref{eq:plastic_strain} and the full stress-strain curve using $\tau = G (\gamma - \gamma_{\rm p})$. 
The yield point is then determined from the condition $\partial \gamma_{\rm p} / \partial \gamma = 1$ for each of these curves.
This yields a distribution of stresses for each strain value, as well as distributions for the yield stress~$\tau_{\rm y}$ and the yield strain~$\gamma_{\rm y}$. 

\subsection*{The microscopic estimates of plasticity}
During an avalanche, the system can be viewed microscopically as comprising a core region of plastically deformed particles, while the surrounding material undergoes elastic deformation in response to the stress induced by the applied strain. A particle is classified as “plastic” if its displacement exceeds 0.30~\AA. This displacement threshold is chosen to include particles actively involved in plastic deformation while excluding those responding elastically.

The fraction of particles participating in the plastic response during an avalanche is quantified as: $n_{\rm{p}} = N_{\rm{p}}/N$, where $N_{\rm{p}}$ is the number of plastically displaced particles, and $N$ is the total number of particles. To perform a comparative study with the bulk plastic strain, we also observe the particles cumulatively ($n_{\rm c,p}$) over a range of strain values. Additionally, a pair of such particles is considered part of the same ``cluster" if they lie within the first coordination shell, as defined by the pair correlation function $(g(r))$. A cluster is considered percolating if it spans the entire system and its size scales with the system size.

\section*{Acknowledgements}
A.D.S.P., S.B., and M.J.A. are supported by the European Union Horizon 2020 research and innovation program under grant agreement no. 857470 and from the European Regional Development Fund via the Foundation for Polish Science International Research Agenda PLUS program grant No. MAB PLUS/2018/8. 
M.J.A. acknowledges support from the Academy of Finland (361245 and 317464) and from the Finnish Cultural Foundation.
S.B. acknowledges support from the National Science Center in Poland through the SONATA BIS grant DEC-2023/50/E/ST3/00569 and from the Foundation for Polish Science in Poland through the FIRST TEAM FENG.02.02-IP.05-0177/23 project.
T.M. and M.J.A. acknowledge support from the FinnCERES ﬂagship (151830423), Business Finland (211835, 211909, and 211989), the Research Council of Finland (13361245), and Future Makers programs.
The authors acknowledge the computational resources provided by the Aalto University School of Science “Science-IT” project.

%\bibliography{biblio}

\end{document}

\renewcommand{\thefigure}{S\arabic{figure}}

\title{Supporting Information: Growth and prediction of plastic strain in metallic glasses}
\author{Tero Mäkinen$^{a,*1}$, Anshul D. S. Parmar$^{b,*2}$, Silvia Bonfanti$^{b,c}$, Mikko Alava$^{a,b}$}
\affiliation{
    $^a$ Aalto University, Department of Applied Physics, PO Box 11000, 00076 Aalto, Espoo, Finland\\
    $^b$ NOMATEN Centre of Excellence, National Center for Nuclear Research, ul. A. Soltana 7, 05-400 Swierk/Otwock, Poland\\
    $^c$ Center for Complexity and Biosystems, Department of Physics `Aldo Pontremoli', University of Milan, Milano, Italy
}
\email{T.M. and A.D.S.P contributed equally to this work.
To whom correspondence should be addressed. E-mails: $^1$tero.j.makinen@aalto.fi and $^2$anshul.parmar@ncbj.gov.pl}

\begin{abstract}
In the supporting information, we provide additional
information regarding the following aspects: 
(i)~System size effects, 
(ii)~{Evolution of the shear modulus},
(iii)~{Alternative yield strain definition},
(iv)~{Posterior distributions},
(v)~{Posterior distributions of the model parameters},
(vi)~{Prefactor of the plasticity model}, 
(vii)~{Particle’s irreversible rearrangement under deformation and the fractal dimension}, and 
(viii)~{Extension to different compositions}.
\end{abstract}

\maketitle

\section*{System size effects}
\begin{figure*}[p!]
    \centering
    \includegraphics[width=0.66\textwidth]{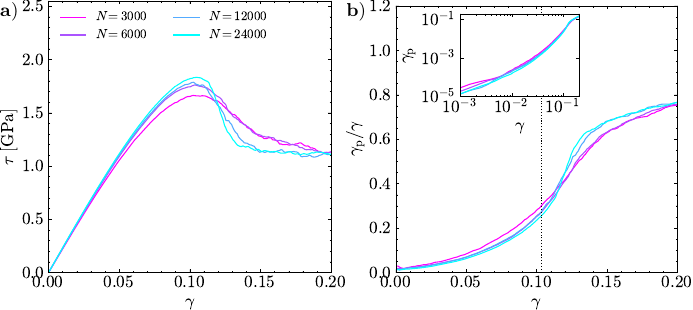}
    \caption{\textbf{a})~The stress-strain curves for different system sizes~$N$ with the same cooling rate $\dot{T} = 10^{12}$~K/s, showing the minimal difference pre-yield above $N=3000$.
    \textbf{b})~The proportion of plastic strain as a function of the total strain, extracted from the stress-strain curves in panel a (and with the same color code). The dashed vertical line illustrates the yield strain averaged over all system sizes. The inset shows the plastic strain in a log-log plot.}
    \label{fig:N}
\end{figure*}

As the system size used in the main text, $N=6000$, is fairly small, we ran simulations on larger---ranging up to $N=24000$---and smaller~($N=3000$) systems. 
The cooling rate is kept constant $\dot{T} = 10^{12}$~K/s, and 100 realizations are used for $N \leq 6000$ and 50 realizations for $N > 6000$.
The resulting stress-strain curves (Fig.~\ref{fig:N}a) and plastic strain evolution (Fig.~\ref{fig:N}b) show that while the behavior of $N=3000$ differs slightly from the from the others, the pre-yield plasticity behaves almost exactly similarly in systems of size $N \geq 6000$.

\section*{Evolution of the shear modulus}
\begin{figure*}[p!]
    \centering
    \includegraphics[width=\columnwidth]{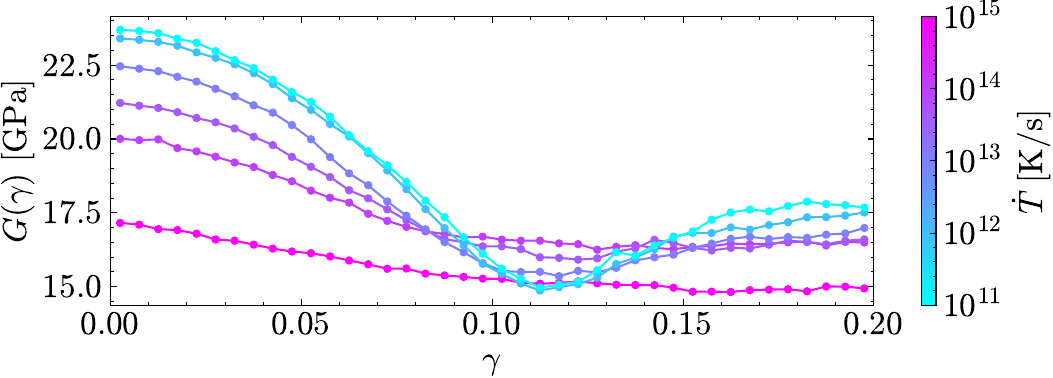}
    \caption{The evolution of the modulus $G(\gamma)$ with strain $\gamma$. The value at the lowest $\gamma$ (determined between $\gamma = 0$ and $\gamma = 0.5$~\%) is the constant value used in the main paper.}
    \label{fig:G}
\end{figure*}

In the methodology outlined in the main paper, the shear modulus $G$ is taken to be constant, and determined from the initial part of the stress-strain curves. However, the shear modulus is known to vary during the deformation 
and we have shown this variation as a function of strain in Fig.~\ref{fig:G}. The modulus $G(\gamma)$ is extracted using the same method as in the main paper, but for consecutive strain itervals of 0.5~\%. One can see that the modulus starts from a value highly determined by the cooling rate, but decreases to a roughly constant value at yielding, independent of the cooling rate.

The change in the modulus is smallest at high cooling rates, where we see the percolation-like power-law plastic strain increase dominate. This means that in the other limiting case---slow cooling rates with an exponential plastic strain accumulation---the departure from the elastic $\tau = G \gamma$ behavior with a fixed $G$ actually has contributions from both the change in the elastic properties (change in $G$) and from the plastic strain accumulation via avalanches~\cite{makinen2025avalanches}.

However, this has no effect on our prediction scheme. For the reconstruction of the stress-strain behavior, it does not matter if one includes both the evolution of $G$ and $\gamma_{\rm p}$ or combines the two effects into the evolution of $\gamma_{\rm p}$. From a practical standpoint, the separation of a change in $G$ and a change in $\gamma_{\rm p}$ is usually not possible e.g.~in experimental conditions. Thus, the standard practice is to take the modulus to correspond to the initial slope of the stress-strain curve. One then defines the plastic strain as the deviation from the elastic behavior given by this modulus, as we have done in this work.

\section*{Alternative yield strain definition}
\begin{figure*}[p!]
    \centering
    \includegraphics[width=\columnwidth]{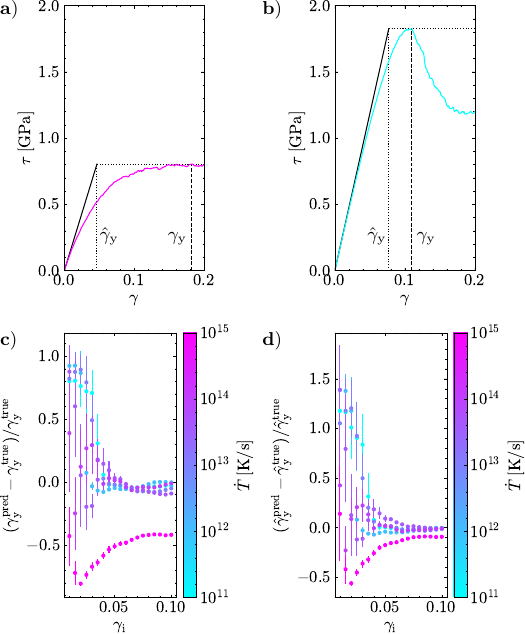}
    \caption{Illustration of the two different yield strain definitions for \textbf{a})~the fastest cooling rate $\dot{T} = 10^{15}$~K/s and \textbf{b})~the slowest cooling rate $\dot{T} = 10^{11}$~K/s. The vertical lines indicate the two yield strains $\gamma_{\rm y}$ and $\hat{\gamma}_{\rm y}$. The relative differences between the predicted $\gamma_{\rm y}^{\rm pred}$ and true $\gamma_{\rm y}^{\rm true}$ yield strains are shown for \textbf{c})~$\gamma_{\rm y}$ and \textbf{d})~the alternative $\hat{\gamma}_{\rm y}$, for all cooling rates and prediction strains $\gamma_{\rm i}$.}
    \label{fig:altgamma}
\end{figure*}

In the main paper we define the yield point $(\gamma_{\rm y}, \tau_{\rm y})$ through the maximum of the stress strain curve. An alternative way is to define the yield strain as the the elastic strain at the maximum stress
$\hat{\gamma}_{\rm y} = \tau_{\rm y} / G$. 
These definitions are illustrated in Fig.~\ref{fig:altgamma} for the fastest cooling rate with an ill-defined yield point (Fig.~\ref{fig:altgamma}a) as well as for the slowest cooling rate (Fig.~\ref{fig:altgamma}b).

Predicting the yield strain using the original definition results in bad performance with the fastest cooling rate (Fig.~\ref{fig:altgamma}c), being around 50~\% off at $\gamma_{\rm i} = 0.05$.
However, using the alternative yield strain definition $\hat{\gamma}_{\rm y}$ (Fig.~\ref{fig:altgamma}d) results in predictions that are within 5~\% for all cooling rates.

\section*{Posterior distributions}
\begin{figure*}[p!]
    \centering
    \includegraphics[width=\columnwidth]{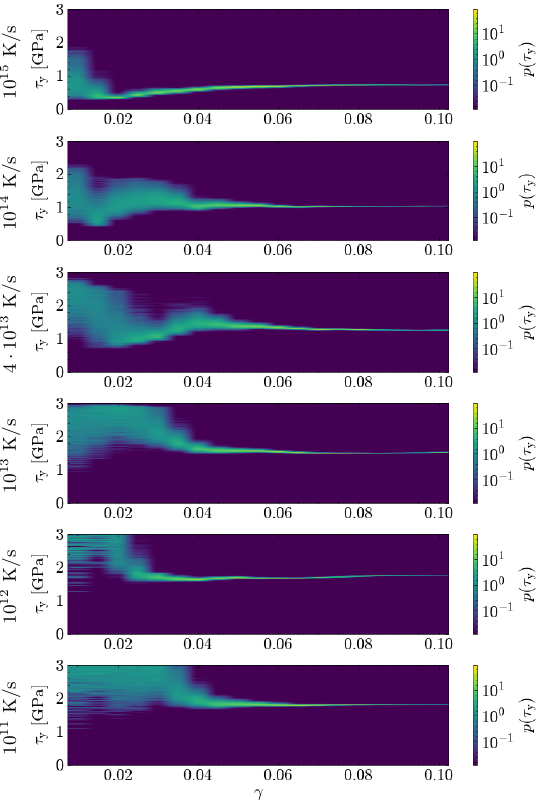}
    \caption{The prediction (posterior distribution) of the yield stress $\tau_{\rm y}$ at different prediction strains $\gamma_i$ for all the cooling rates $\dot{T}$.}
    \label{fig:tauy}
\end{figure*}

In the main paper, Fig.~2 shows the posterior distribution of the yield stress predictions $p(\tau_{\rm y})$ for $\gamma_{\rm i}=5$~\%, as well as the convergence of the mean and standard deviation of this prediction for increasing $\gamma_{\rm i}$.
Here, Fig.~\ref{fig:tauy} shows the full distribution for all the values of $\gamma_{\rm i}$ from 1~\% to 10~\%. One can clearly see the extremely wide distribution for low values of $\gamma_{\rm i}$ and the convergence to a narrow distribution around $\gamma_{\rm i} = 5$~\%.

\begin{figure*}[p!]
    \centering
    \includegraphics[width=\columnwidth]{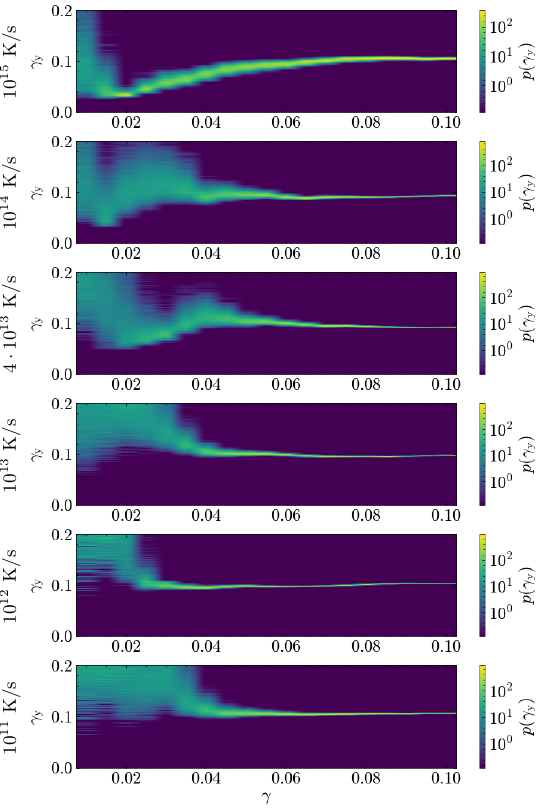}
    \caption{The prediction (posterior distribution) of the yield strain $\gamma_{\rm y}$ at different prediction strains $\gamma_i$ for all the cooling rates $\dot{T}$.}
    \label{fig:gammay}
\end{figure*}

The exact same is true for the posterior distribution of the yield strain predictions $p(\gamma_{\rm y})$ (Fig.~\ref{fig:gammay}), but for cooling rates $\dot{T} > 10^{13}$~K/s the convergence happens slightly later, around $\gamma_{\rm i} = 6$~\%. With the fastest cooling rate ($\dot{T} = 10^{15}$~K/s) one can see convergence of the distribution really early, at around $\gamma_{\rm i} = 2$~\%, but with the mean of this narrow distribution increasing with $\gamma_{\rm i}$.
This is related to the different yielding mode discussed in the main text.

\section*{Posterior distributions of the model parameters}
\begin{figure*}[p!]
    \centering
    \includegraphics[width=\columnwidth]{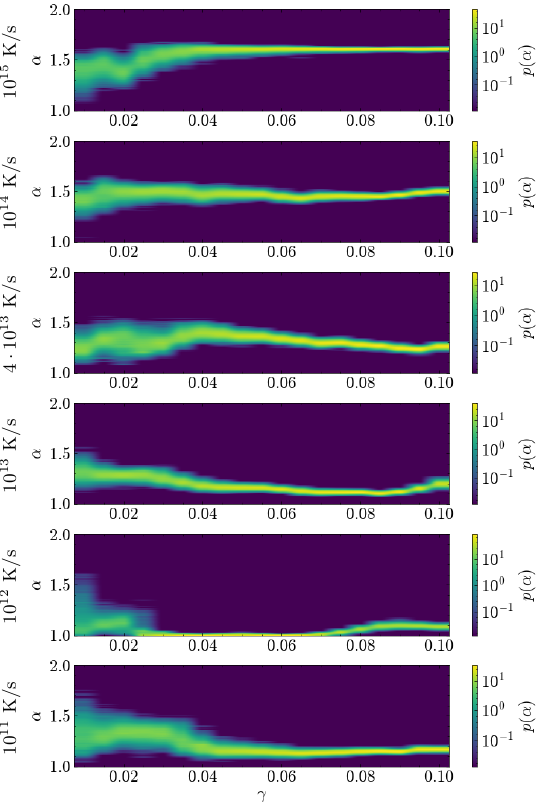}
    \caption{The posterior distribution of the $\alpha$ parameter of the plasticity model at different prediction strains $\gamma_{\rm i}$ for all the cooling rates $\dot{T}$.}
    \label{fig:c}
\end{figure*}

The posterior distributions of $\tau_{\rm y}$ and $\gamma_{\rm y}$ result from the posterior distributions of the parameters of the underlaying plasticity model (Eq.~1 of the main text), namely $\alpha$ and $\gamma_0$. Similarly to Figs.~\ref{fig:tauy} and~\ref{fig:gammay}, Fig.~\ref{fig:c} shows the evolution of the posterior distribution of the exponent $\alpha$ of the plasticity model with the prediction strain $\gamma_{\rm i}$. For the fastest cooling rates, dominated by the power-law increase in plastic strain, the distribution $p(\alpha)$ quickly converges to a narrow one. As the cooling rate decreases, and the exponential behavior starts to dominate, some fluctuations in the posterior means can be seen.

\begin{figure*}[p!]
    \centering
    \includegraphics[width=\columnwidth]{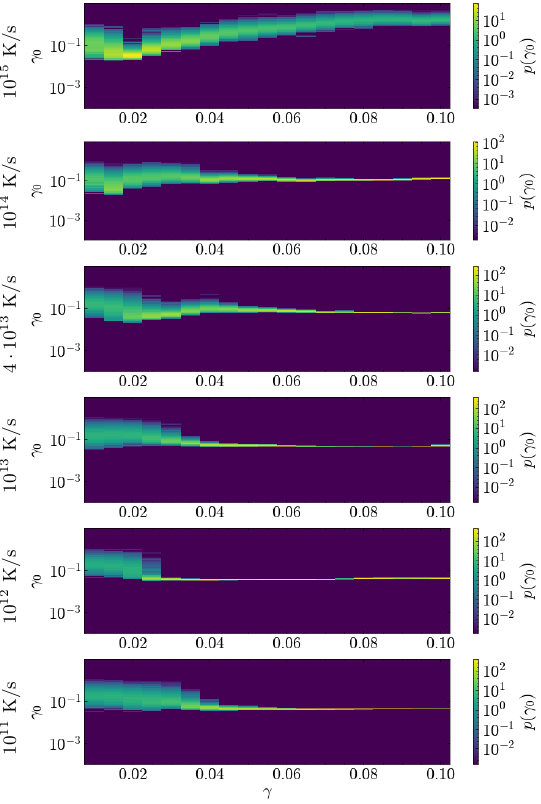}
    \caption{The posterior distribution of the $\gamma_0$ parameter of the plasticity model at different prediction strains $\gamma_{\rm i}$ for all the cooling rates $\dot{T}$.}
    \label{fig:gamma0}
\end{figure*}

This switch from power-law dominated behavior to exponential behavior is seen in the evolution of the posterior distribution of the characteristic strain of the exponential term~$\gamma_0$ (Fig.~\ref{fig:gamma0}). At slow cooling rates, $p(\gamma_0)$ quickly converges to an extremely narrow distribution with a low mean value. For the fast cooling rates there are more fluctuations in the mean and the distribution is wider. This is understandable, as high values of $\gamma_0$ make the exponential term $e^{\gamma/\gamma_0}$ roughly constant at low strains, thus making the exact determination of $\gamma_0$ hard and unnecessary for plasticity predictions.

\section*{Prefactor of the plasticity model}
\begin{figure*}[p!]
    \centering
    \includegraphics[width=0.5\textwidth]{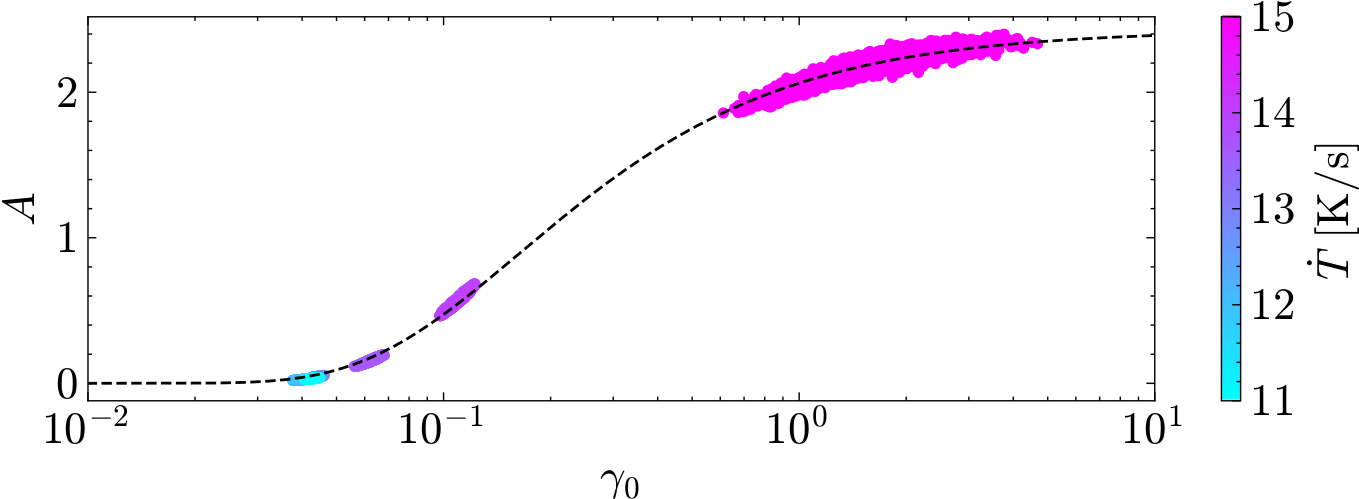}
    \caption{Samples of the prefactor $A$ of the plasticity model as a function of the exponential scale parameter $\gamma_0$, for all the different cooling rates, inferred at $\gamma_{\rm i} = 0.08$. The dashed black line shows a fit $A \propto e^{- C/\gamma_0}$, illustrating that $A$ is directly linked to $\gamma_0$.}
    \label{fig:A}
\end{figure*}

We have not discussed the prefactor $A$ of the plasticity model in detail, as it is directly linked to the other parameters, namely the characteristic strain of the exponential term $\gamma_0$. Doing the Bayesian inference for all the cooling rates at $\gamma_{\rm i} = 0.08$ and sampling from the posterior distributions of $A$ and $\gamma_0$ gives the points shown in Fig.~\ref{fig:A}. One can clearly see that the points lie on a path given by a master curve, directly expressing $A$ as a function of $\gamma_0$. As a guide to the eye, we have fit $A \propto e^{-C/\gamma_0}$ to these points, illustrating this link more concretely.

\section*{Particle's irreversible rearrangement under deformation and the fractal dimension}
\begin{figure*}[p!]
    \centering
    \includegraphics[width=0.5\textwidth]{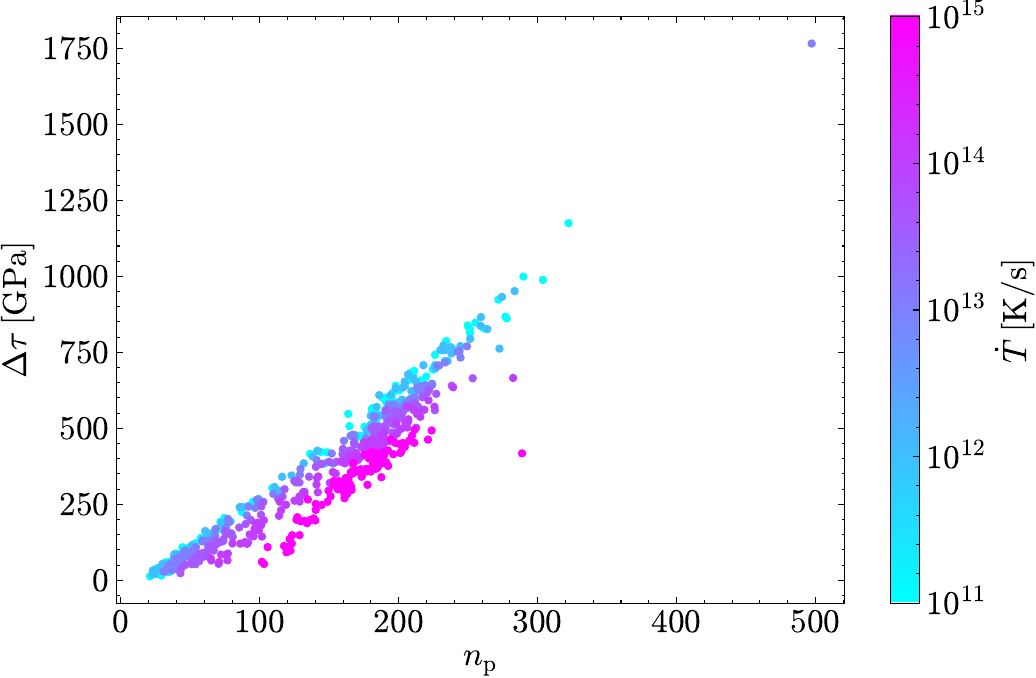}
    \caption{The fraction of particles involved in local plastic rearrangements ($n_{\text{p}}$) and the macroscopic stress drop ($\Delta \tau$) during individual plastic events, shown for a range of cooling rates. The linear behaviour suggests that $n_{\text{p}}$ effectively captures the local plasticity and reflects to the system’s collective response during stress release.}
    \label{fig:npl_vs_Ds}
\end{figure*}

In addition to the microscopic description of plasticity and the participation ratio of all particles during an avalanche~\cite{makinen2025avalanches}, we examined the correlation between the fraction of particles involved in local rearrangements ($n_{\text{p}}$) and the macroscopic stress drop ($\Delta \tau$) during plastic events. Fig.~\ref{fig:npl_vs_Ds} shows a strong correlation between $n_{\text{p}}$ and $\Delta \tau$ across a range of cooling rates, suggesting that the estimated fraction of participating particles satisfactorily captures the extent of local plasticity.

\begin{figure*}[p!]
    \centering
    \includegraphics[width=0.5\textwidth]{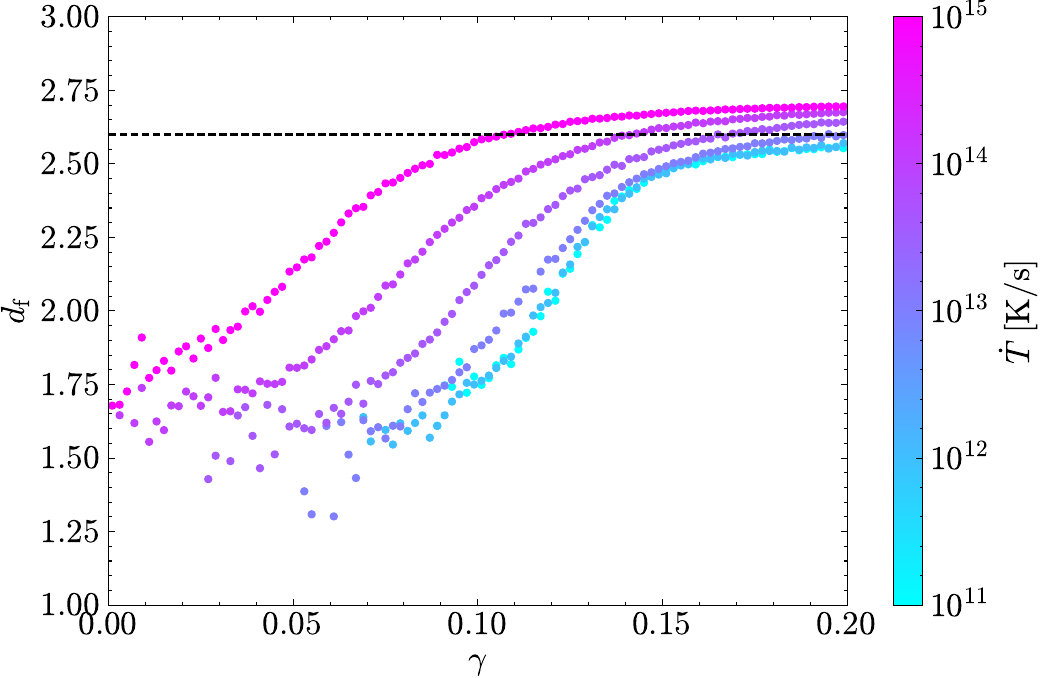}
    \caption{Fractal dimension of the largest cluster of plastically deformed particles for $N=6000$ particle system, prepared at various cooling rates. For poorly annealed glasses, the fractal dimension in the steady-state regime approaches $d_f \sim 2.6$ (dashed line), consistent with the exponent from an isotropic, percolation-like growth of plastic activity in the deep post-yield state.}
    \label{fig:fractal_dim}
\end{figure*}

To support the argument regarding the development and spatial growth of plasticity in "bad" glasses, we estimate the fractal dimension of the largest cluster of plastically deformed particles using the box-counting method. Fig.~\ref{fig:fractal_dim} shows the fractal dimension across a range of cooling rates in the steady-state regime. For poorly annealed glasses, the plastic clusters exhibit a fractal dimension of approximately $d_{\rm f} \sim 2.6$, consistent with an isotropic percolation-like structure in the deep steady state, far beyond the yield point.

\section*{Extension to different compositions}
\begin{figure*}[p!]
    \centering
    \includegraphics[width=\textwidth]{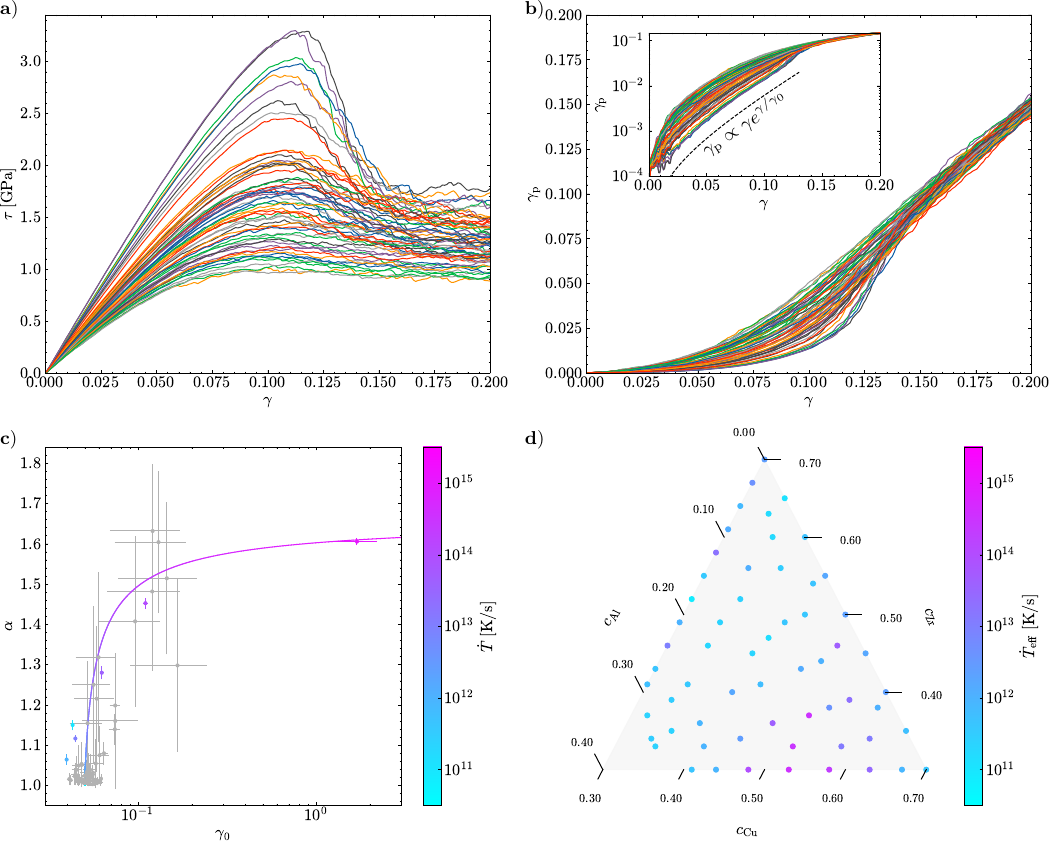}
    \caption{\textbf{a})~The stress-strain curves for different CuZrAl compositions ($\dot{T} = 10^{12}$~K/s) from Ref.~\cite{makinen2025bayesian}.
    \textbf{b})~The proportion of plastic strain for all these compositions. The inset shows the same curves on semi-logarithmic plot to illustrate the $\gamma_{\rm p} \propto \gamma e^{\gamma/\gamma_0}$ behavior (dashed line).
    \textbf{c})~The means and standard deviations of the model parameters $\alpha$ and $\gamma_0$ (inferred at $\gamma_{\rm i} = 8$~\%) for all the CuZrAl compositions (gray), the CuZr systems cooled at different rates (points colored according to the cooling rate), as well as the behavior given by the fits of Fig.~3 of the main paper (line colored according to the cooling rate).
    \textbf{d})~The effective cooling rate $\dot{T}_{\rm eff}$ for all the CuZrAl compositions.
    }
    \label{fig:allglasses}
\end{figure*}

The main paper discusses only data from equiatomic CuZr systems with different cooling rates. By including the data from Ref.~\cite{makinen2025bayesian} for a large number of simulations on a non-equiatomic CuZrAl system at a single cooling rate (here $\dot{T} = 10^{12}$~K/s) allows for the study of the plasticity behavior as a function of the composition. The stress-strain curves (Fig.~\ref{fig:allglasses}a) show widely varying mechanical response for the different compositions, and Fig.~\ref{fig:allglasses}b illustrates the clear exponential behavior of some compositions, as well as a non-exponential behavior of others. None of the compositions considered at $\dot{T} = 10^{12}$~K/s show the power-law increase of plastic strain observed with equiatomic CuZr at $\dot{T} = 10^{15}$~K/s.

One can however fit Eq.~1 of the main paper (at $\gamma_{\rm i} = 8$~\%) to each of these curves and observe the relation between the model parameters $\alpha$ and $\gamma_0$. The posterior means and standard deviations are drawn in grey in Fig.~\ref{fig:allglasses}c. Including the fit (also at $\gamma_{\rm i} = 8$~\%) of the equiatomic CuZr at different cooling rates (points colored according to the cooling rate in Fig.~\ref{fig:allglasses}c), as well as the relation given by the fits shown in Fig.~3 of the main paper (colored line in Fig.~\ref{fig:allglasses}c), shows that the CuZrAl fits are not far from the curve. The fits of Fig.~3 of the main paper are a logistic fit (between $1$ and $5/3$) for $\alpha$ as a function of $\log \dot{T}$ (with a transition at $\dot{T} = 10^{13.71}$~K/s, and an exponential fit $\gamma_0 = \gamma_0^\infty e^{\dot{T} / \dot{T}_{\rm trans}}$ with a characteristic value of $\dot{T}_{\rm trans} = 2.78 \cdot 10^{14}$~K/s and a slow cooling limit $\gamma_0^\infty = 4.89 \cdot 10^{-2}$. 

This motivates the determination of an effective cooling rate $\dot{T}_{\rm eff}$ by finding the closest point on the curve for each of the CuZrAl points.
One can then plot these effective cooling rates in a ternary diagram (Fig.~\ref{fig:allglasses}d) to illustrate the change in this quantity with composition. This qualitatively lines up with the results of Ref.~\cite{makinen2025bayesian} as the highest $\dot{T}_{\rm eff}$ are in the region showing the worst mechanical properties, and the lowest $\dot{T}_{\rm eff}$ are close to the region exhibiting the best mechanical properties.


%\bibliography{biblio}